\documentclass[10pt,twocolumn,letterpaper]{article}
\pdfoutput=1

\usepackage[utf8]{inputenc}
\usepackage{graphicx}
\usepackage{cite}

\usepackage{physics}
\usepackage{upgreek}
\usepackage{color}
\usepackage{amssymb}
\usepackage{gensymb}
\usepackage{floatrow}
\usepackage{authblk}
\usepackage{titlesec}
\renewcommand{\thesection}{\Roman{section}}
\titleformat{\section}{\Large\scshape}{\thesection}{0.7em}{}

\usepackage[
 format=plain, 
 labelsep=space,
 labelfont=bf, 
 ]{caption}

\begin{document}

\title{Spatial Control of Hybridization-Induced Spin-Wave Transmission Stop Band} 
\author[1]{Franz Vilsmeier\footnote{franz.vilsmeier@tum.de}}

\author[1]{Christian Riedel}
\author[1]{Christian H. Back}
\affil[1]{\textit{Fakultät für Physik, Technische Universität München, Garching, Germany}}
\date{March 23, 2024}

\maketitle

\begin{abstract}
\noindent Spin-wave (SW) propagation close to the hybridization-induced transmission stop band is investigated within a trapezoid-shaped 200\,nm thick yttrium iron garnet (YIG) film using time-resolved magneto-optic Kerr effect (TR-MOKE) microscopy and broadband spin wave spectroscopy, supported by micromagnetic simulations. The gradual reduction of the effective field within the structure leads to local variations of the SW dispersion relation and results in a SW hybridization at a fixed position in the trapezoid where the propagation vanishes since the SW group velocity approaches zero. By tuning external field or frequency, spatial control of the spatial stop band position and spin-wave propagation is demonstrated and utilized to gain transmission control over several microstrip lines.
\end{abstract} 

\bigskip

\section{Introduction}
Driven by potential spin-wave-based applications in computing and data processing, the field of magnonics has garnered growing interest in recent years~\cite{Lee2008,Schneider2008,VVKruglyak2010,Khitun2011,Sadovnikov2015,Chumak2015,Wang2018,Wang2020,Barman2021,Papp2021,Chumak2022}. To perform logic operations encoded within magnon currents various approaches were suggested and realized, such as interference-based logic gates~\cite{Lee2008,Goto2019,Papp2021,Qin2021} or magnonic crystals that exploit the periodicity-induced formation of bandgaps in the spin-wave spectrum~\cite{Gubbiotti2010,Serga2010,Lenk2011,Wang2010,Krawczyk2014,Morozova2015,Kreil2019,Lisiecki2019}. These devices rely on precise control and manipulation of spin-waves with wave vector $\textbf{k}$ within a material with magnetization \textbf{M}. Recently, a hybridization-induced spin-wave-transmission stop band was demonstrated in 200\,nm yttrium iron garnet (YIG)~\cite{Riedel2023}, adding to the list of options for engineering spin-wave propagation. It was shown that the hybridization of two different Damon Eshbach-like (DE) ($\textbf{k}\perp\textbf{M}$) SW modes causes a frequency- and field-dependent suppression of SW propagation in a film with in-plane magnetization. Furthermore, it is well known that at the edges of thin magnetic films, depending on the magnetization direction, the effective field is locally reduced in order to avoid the generation of magnetic surface charges~\cite{Coey2001,Lenk2011,Gruszecki2014}. This allows for shape-modulated local variations of SW propagation. Combining this effect with the transmission stop band may provide enhanced control over spin-wave propagation dynamics and facilitate the implementation of magnonic devices.

\begin{figure*}[t]
    \includegraphics[width=1\textwidth]{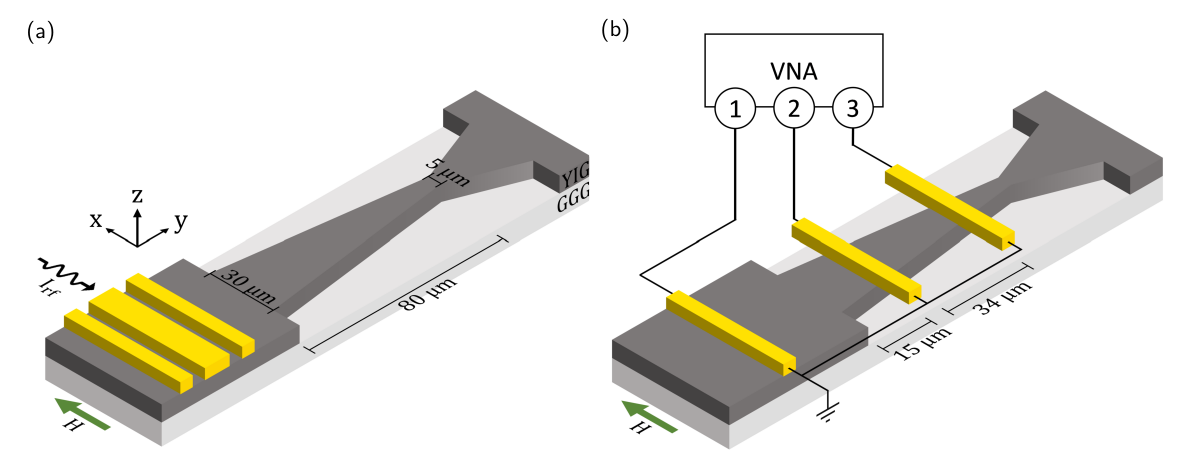}
    \caption{Sketch of the experimental setup. (a) Schematic for TR-MOKE measurement. Spin-waves are excited in the dipolar regime by the CPW and propagate through the trapezoid structure. The trapezoid was chosen to have a maximum width of 30\,$\upmu$m, a minimum width of 5\,$\upmu$m and a length of 80\,$\upmu$m. A static external field along the $x$-direction was applied throughout the experiment. (b) Schematic of all-electrical VNA spin-wave spectroscopy measurement. Spin-waves are excited from the first microstrip and detected via two more microstrips at different positions along the trapezoid geometry. Each microstrip is connected to a separate port of a four-port VNA.}
    \label{fig:Figure1} 
\end{figure*}

In this report, we investigate the effect of the geometry-induced variation of the effective field in a 200\,nm YIG film on the hybridization-induced stop band. We demonstrate that the spin-wave propagation distance can be actively controlled within a trapezoid-shaped magnetic film as the reduced effective field locally enforces the hybridization condition. Experimental dispersion measurements and micromagnetic simulations using \textsc{TetraX}~\cite{TetraX2022} are conducted to determine the full film stop band condition. From further micromagnetic simulations using \textsc{MuMax3}~\cite{MuMax2014}, the effective internal field of a trapezoid geometry is determined. An inhomogeneous field distribution with a gradual decrease along the trapezoid's length is observed. We experimentally investigate the corresponding spin-wave propagation within the trapezoid in a DE-like geometry using time-resolved magneto-optic Kerr effect (TR-MOKE) microscopy~\cite{Perzlmaier2008,Farle2012,Au2011,Bauer2014,Stigloher2018} and broadband spin-wave spectroscopy. Bending of wavefronts, the formation of edge channels, and a gradual decrease of wavelength along the propagation direction are observed. At a distinct position in the trapezoid, the propagation ceases. We show that this stop position is locally induced by the reduced effective field, which grants access to the spin-wave transmission stop band. Based on these findings, we demonstrate spatial control of spin-wave propagation within a trapezoid-shaped device by tuning the static external field close to the stop band. We utilize this effect for the active transmission control between microstrip lines.

\section{Experimental Results}

The first set of experiments was carried out using time-resolved magneto-optic Kerr effect (TR-MOKE) microscopy. Here, the dynamic out-of-plane magnetization component $\delta m_z$ is spatially mapped in the $xy$-plane, and a direct observation of spin-wave propagation in the sample is obtained. Simultaneously, the reflectivity is detected, providing a topographic map of the sample. The measurements were conducted on a 200\,nm thick yttrium iron garnet (YIG) film grown by liquid phase epitaxy on a gadolinium gallium garnet (GGG) substrate. The trapezoid shape, with a gradual continuation back to the full film, was patterned by means of optical lithography and subsequent Argon sputtering of the YIG film. For the excitation of spin-waves, a coplanar waveguide (CPW) was fabricated on top of the YIG film by optical lithography and electron beam evaporation of Ti(5\,nm)/Au(210\,nm). During the measurements, the external bias field was fixed along the CPW, so spin-waves in a DE-like geometry were excited~\cite{Damon1961}. A schematic of the measurement geometry can be found in Fig.~\ref{fig:Figure1}(a).

\begin{figure}[h!]
    \includegraphics[width=\textwidth]{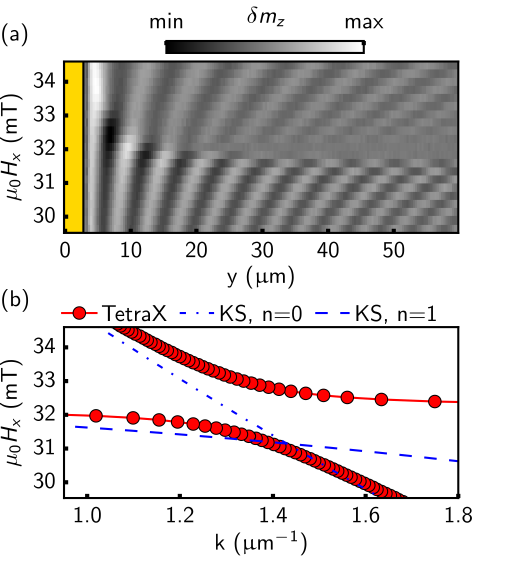}
    \caption{(a) Measurement of SW propagation excited by the CPW (gold) in full film YIG as a function of the external field. A suppression of propagation is visible around 32\,mT. The grey-scale represents the measured Kerr amplitude. (b) Micromagnetic simulations with \textsc{TetraX} for $f=2.8$\,GHz. The DE-mode (blue dash-dotted line) and the n=1 mode (blue dashed line) hybridize and form an anticrossing in the micromagnetic simulations (red line). This results in an attenuation of SW propagation since the group velocity approaches zero. For all the simulations, the following material parameters were used: saturation magnetization $M_\text{s}=1.4\cdot10^5\,\frac{\text{A}}{\text{m}}$, exchange stiffness $A_\text{ex}=3.7\cdot10^{-12}\,\frac{\text{J}}{\text{m}}$, gyromagnetic ratio $\gamma=176\,\frac{\text{GHz}}{\text{T}}$, film thickness $L=200$\,nm.}
    \label{fig:Figure2}
\end{figure}

As a preliminary step, the spin-wave stop band in the unpatterned plane YIG film was identified by examining SW propagation far away from the patterned trapezoid structure. In this context, line scans of the Kerr signal along the $y$-direction were recorded as a function of the applied external field at a constant microwave frequency of $f=2.8$\,GHz. The result is depicted in Fig.~\ref{fig:Figure2}(a). Here, a clear suppression of spin-wave propagation around 32\,mT can be observed. Previous work~\cite{Vaatka2021,Riedel2023} has shown that hybridization between the DE-mode and the first-order perpendicular standing spin-wave (PSSW) mode can create a spin-wave stop band in 200\,nm YIG. This is further illustrated in Fig.~\ref{fig:Figure2}(b) by micromagnetic simulations with \textsc{TetraX}~\cite{TetraX2022}, an open-source Python package for finite-element-method micromagnetic modelling~\cite{TetraX2022}. In zeroth-order perturbation theory, according to Kalinikos and Slavin (KS)~\cite{Kalinikos1986}, the n=0 mode (blue dash-dotted line) and n=1 mode (blue dashed line) cross each other. This degeneracy is lifted by the formation of an avoided crossing in the micromagnetic simulations (red line). This leads to a flattening of the dispersion relation and, in turn, to a decrease in group velocity~\cite{Riedel2023}. For the given experimental parameters, the stop band is predicted at approximately 32\,mT, consistent with the observed suppression of propagation in the full film line scans. We thus conclude that the pronounced attenuation can be attributed to the hybridization-induced stop band.

\begin{figure}[t]
    \includegraphics[width=\textwidth]{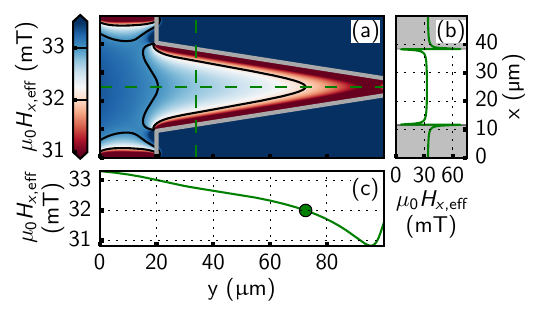}
    \caption{Micromagnetic simulations of the $x$-component of the effective field inside the trapezoid structure. The effective magnetic field varies locally across the geometry in (a). In the vicinity of edges, it is significantly reduced. The grey contour depicts the spatial boundaries of the simulated YIG structure. Across the width of the structure shown in (b), a strong dip of the field at the edges of the geometry is visible. Here, the grey-shaded rectangles indicate the areas outside of the magnetic structure. The effective field along the length of the trapezoid gradually decreases, as shown in (c). The full film hybridization condition at 2.8\,GHz is marked with a green dot.}
    \label{fig:Figure3}
\end{figure}

To understand the influence of the trapezoid geometry on SW propagation, and in turn, on the hybridization condition, further micromagnetic simulations were performed~\cite{MuMax2014} to determine the effective field of the tapered SW waveguide. Fig.~\ref{fig:Figure3}(a) shows the spatial distribution of the $x$-component $\mu_0H_{x,\text{eff}}$ of the effective field at an externally applied field $\mu_0H_x=33.5$\,mT. The simulations revealed that the effective field varies locally inside the trapezoid and is strongly reduced at the YIG edges. The iso-field lines (black lines), which display rounded triangular-like features, further illustrate the inhomogeneous spatial distribution.
Along the width of the trapezoid (Fig.~\ref{fig:Figure3}(b)), we observe sharp edge pockets of low internal field, and a gradual decrease of field along the axis of spin-wave propagation (Fig.~\ref{fig:Figure3}(c)). The origin of this inhomogeneity of the effective field lies in the geometry-induced demagnetizing field, which aims to avoid the formation of magnetic surface charges~\cite{Coey2001,Gruszecki2014}. 

Another important consideration regarding the effect of the modified trapezoid waveguide is the emergence of additional width modes in the SW dispersion relation due to the finite waveguide width~\cite{Demidov2008,Xing2013,Chumak2019}. However, this width quantization does not affect the PSSWs, and the intersection of modes is still present. Thus, we argue that the key influence of the trapezoid geometry on the stop band is the reduction in effective field which results in a local variation of the dispersion relation. A more detailed discussion concerning the width quantization can be found in the supplementary material.

\begin{figure}[hbt!]
    \includegraphics[width=\textwidth]{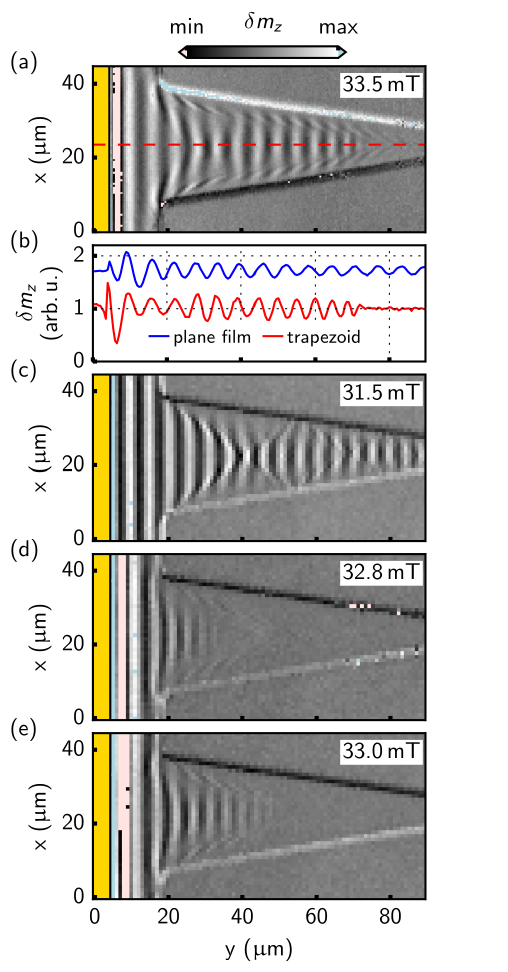}
    \caption{Kerr images at 2.8\,GHz. (a) Propagation of the main mode stops at a distinct position in space. The golden area depicts the excitation source. Light red and blue indicate saturation of the grey-scale. (b) Gradual decrease in wavelength and suppression of propagation along the trapezoid (red line in (a)) is observed. Propagation in the non-patterned plane YIG (blue curve) is also shown. (c) Below the hybridization field, propagation along the full trapezoid is observed. (d)-(e) By slightly tuning the field above the stop band, spin-wave propagation vanishes at different positions in space.}
    \label{fig:Figure4}
\end{figure}

Next, we experimentally investigate the effect of the geometry-induced field distribution on spin-wave propagation. Fig.~\ref{fig:Figure4}(a) displays TR-MOKE measurements in which plane spin-waves are launched from the CPW into the trapezoid in the $y$-direction, with an excitation frequency $f=2.8$\,GHz and a static external field $\mu_0H_x=33.5$\,mT. Changes in the propagation characteristics are observed upon entering the trapezoid. Apart from a prominent mode with slightly bent wavefronts in the trapezoid center, a localized mode with strongly bent wavefronts close to the edges appears. We also note that a magnetic contrast right at the edges of the patterned structure was observed in some Kerr images. We argue that this artifact is due to imperfections in the fabrication process and discuss it in more detail in the supplementary material. 

The observed bending of wavefronts can be attributed to the inhomogeneous internal field profile \cite{Gruszecki2014}, where a local reduction in the effective field causes a shift towards lower fields and lower wavelengths in the spin-wave dispersion relation. Additionally, the edge localization of modes is a direct consequence of the low-field pockets in the effective field distribution (Fig.~\ref{fig:Figure3}(b)) as reported previously~\cite{Topp2008,Jorzick2002,Bauer2014}.

Furthermore, the center mode changes wavelength as it travels through the trapezoid structure. Notably, it comes to a halt at a specific position in space, beyond which the spin-wave propagation is almost entirely suppressed. This behaviour is illustrated in more detail in Fig.~\ref{fig:Figure4}(b), where a $y$-line profile (red curve) along the red dashed line in Fig.~\ref{fig:Figure4}(a) is plotted. A line scan on plane YIG (blue curve), far away from any patterned structure, is also presented for comparison. As the spin-wave enters the trapezoid, its wavelength gradually decreases up to more than 50\%, consistent with the simulated decrease in the effective field (Fig.~\ref{fig:Figure3}(c)). However, the propagation abruptly ceases at a specific position in space ($y\approx74\,\upmu$m). From Fig.~\ref{fig:Figure3}(c), we observe that the stop position of the center mode within the wedge corresponds to an estimated effective field of about 32\,mT which aligns well with the measured full film hybridization condition (Fig.~\ref{fig:Figure2}). Thus, we conclude that the reduction in effective field at different positions in space leads to the local dispersion entering the hybridization regime at a specific position in space, resulting in a sharp local attenuation of spin-wave propagation.

Now, we aim to apply our findings towards the active manipulation of spin-wave propagation. To this end, additional Kerr images as a function of the external field were taken and are depicted in Figs.~\ref{fig:Figure4}(c)-(e). Below the hybridization field, at 31.5\,mT (Fig.~\ref{fig:Figure4}(c)), propagation along the full length of the trapezoid without any sharp attenuation is present. No spatial suppression of propagation is observable since the effective field is only further reduced inside the trapezoid, and thus, the stop band regime is never reached. We also note a complex spatial beating profile with a prominent node at $y\approx45\,\upmu$m, and several less prominent ones. This self-focusing effect results from interference of the width modes induced by the tapered waveguide geometry and has been reported in magnonic microstripes before~\cite{Demidov2008,Xing2013}. Furthermore, caustic-like beams induced by the corners where the full film transitions into the trapezoid may emerge~\cite{Riedel2023,Wartelle2023}. These caustic-like beams are reflected back and forth at the edges, resulting in non-equidistant areas of higher and lower amplitude. 

On tuning the external field slightly above 32\,mT (Figs.~\ref{fig:Figure4}(d)-(e)), however, the spin-wave propagation ceases at different positions in space. Furthermore, the boundaries of the spin-wave pattern display a shape reminiscent of the iso-field lines in the effective field. As the external field increases, the positions where the dispersion relation locally gains access to the transmission stop band also shift further outward along the $y$-direction. As a result, the geometry-induced hybridization allows to actively control the spin-wave propagation distance merely by tuning the external field within a reasonable range. 

\begin{figure*}[t]
    \includegraphics[width=\textwidth]{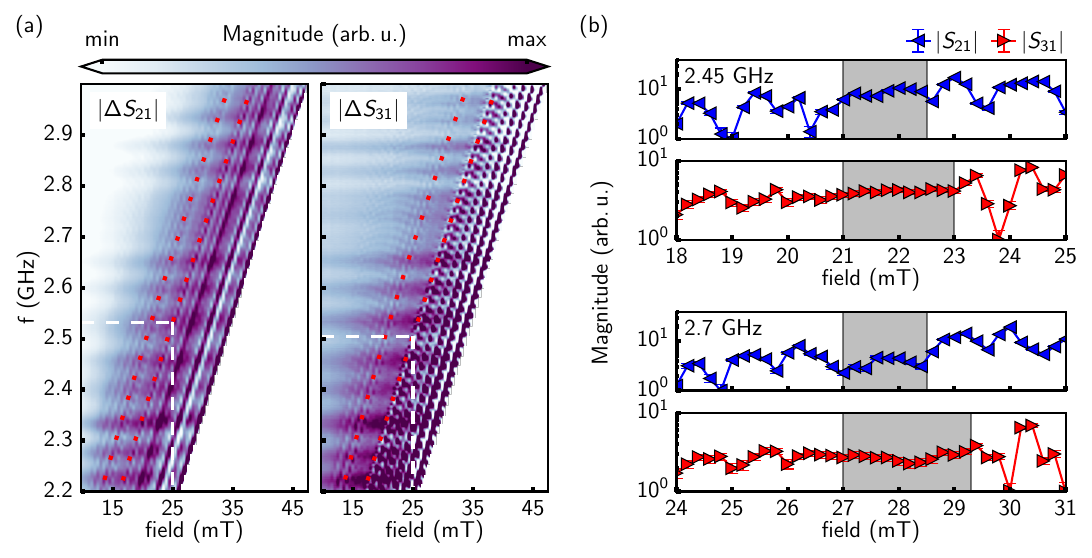}
    \caption{Selective control of transmission between microstrips along trapezoid-like structure. (a) Spin-wave transmission spectra (amplitudes $|\Delta S_\mathrm{21}|$ and $|\Delta S_\mathrm{31}|$). A region of low to no transmission consistent with the expected hybridization conditions occurs in both spectra (red dotted lines serve as guides to the eye). This region appears broader in the transmission spectrum from port 1 to port 3. Moreover, at a given external field, the stop band starts at lower frequencies in $|\Delta S_\mathrm{31}|$ compared to $|\Delta S_\mathrm{21}|$ as highlighted by white dashed lines. (b) Transmission signals in CW mode at fixed frequencies. The hybridization-induced stop band (roughly marked by gray-shaded areas) spans to higher applied fields in the $|S_\mathrm{31}|$ transmission trace.}
    \label{fig:Figure5}
\end{figure*}

Further all-electrical Vector Network Analyzer (VNA) spin-wave spectroscopy measurements were conducted with the intention to demonstrate the control of spin-wave propagation within a potential magnonic device. For this purpose, three 800\,nm wide Au microstrips were patterned at different positions along the trapezoid structure. One microstrip served as a source of spin-wave excitation, while the other two served for detection. The microstrips were connected to separate ports of a four-port VNA, and broadband spin-wave spectroscopy was performed. Note that the choice to employ microstrips instead of CPWs was made in order to obtain a more continuous range of wavenumbers for both the excitation and detection processes. A sketch of the measurement geometry is depicted in Fig.~\ref{fig:Figure1}(b).

Fig.~\ref{fig:Figure5}(a) displays the detected spin-wave transmission spectra showcasing the amplitudes of the scattering parameters $S_\mathrm{21}$, $S_\mathrm{31}$ in terms of $|\Delta S_\mathrm{21}|$ and $|\Delta S_\mathrm{31}|$. We point out that, in the following discussion, $|S_\mathrm{ij}|$ denotes the absolute values of the detected scattering parameters whereas $|\Delta S_\mathrm{ij}|$ refers to absolute values where a high-field subtraction method was applied. Also note that for better visibility, only data close to the stop band condition is depicted. Full transmission spectra, along with more detailed information about the data processing procedure, can be found in the supplementary material. Both spectra exhibit amplitude oscillations with the $|\Delta S_\mathrm{31}|$ spectrum displaying shorter spacing between these oscillations. This is due to the change of the lateral spin wave profile, where the positions of high-amplitude and caustic-like nodes shift due to changes in the external magnetic field and applied frequency. This effect leads to a smaller node spacing in the field domain at the location of the third microstrip due to the gradual decrease in trapezoid width~\cite{Demidov2008}.  

Moreover, distinct wide regions with low to no transmission in the spectra (highlighted by red dotted lines) occur at conditions in accordance with the spin wave stop band. For the transmission $|\Delta S_\mathrm{31}|$, this band is noticeably broader compared to the $|\Delta S_\mathrm{21}|$ spectrum and is reached at lower frequencies at a given field (compare white dashed lines). This is a direct consequence of the spatially varying occurrence of the hybridization condition suppressing the propagation of spin waves over a broader range of fields and frequencies the further they advance along the trapezoid. To put it differently, distinct external field and frequency conditions exist where transmission is absent in both $|\Delta S_\mathrm{21}|$ and $|\Delta S_\mathrm{31}|$, transmission is observed only in $|\Delta S_\mathrm{21}|$, and transmission occurs in both $|\Delta S_\mathrm{21}|$ and $|\Delta S_\mathrm{31}|$. As a result, selective control over the transmission between the microstrips can be achieved by slightly tuning the frequency or the applied bias field. 

This is further illustrated in the continuous wave (CW) mode measurements at fixed frequencies shown in Fig.~\ref{fig:Figure5}(b). The regions of suppressed transmission shift with applied frequency and span over a broader field range in the $|S_\mathrm{31}|$ parameter. The hybridization-induced stop band (highlighted by gray-shaded regions) extends to higher fields due to localized effective field reduction. For instance, at 2.45\,GHz and with a field of 23\,mT, we observe transmission in the $|S_\mathrm{21}|$ channel but minimal transmission in the $|S_\mathrm{31}|$ trace, similar to 2.7\,GHz at 29\,mT. Interestingly, the transmission in $|S_\mathrm{31}|$ also appears to be suppressed for fields slightly below the hybridization field. Additional TR-MOKE data in the supplementary material reveals that this behavior can be attributed to the formation of caustic-like beams that are significantly attenuated upon propagation along the geometry.

To conclude this section, we suggest employing multiple microstrips along the trapezoid geometry for potential logic operation. Moreover, in the supplementary material, we provide further discussion on properties of the hybridization, such as its thickness dependence.

\section{Conclusion}
In conclusion, we demonstrated the feasibility of actively controlling the spin-wave propagation distance by combining the hybridization-induced stop band and a geometry-induced variation of the effective field in 200\,nm YIG within a trapezoid-shaped magnetic film. Experiments and micromagnetic simulations were performed to gain insight into the effect of the trapezoid geometry on the effective field. The results show that the spin-wave transmission stop band locally induces the stop position, allowing for spatial control of spin-wave propagation within a trapezoid-shaped device by tuning the static external field close to the stop band. Using multiple microstrips along the trapezoid, we further demonstrated the feasibility of active transmission control between microstrips by external field and frequency. The proposed method offers a promising approach for further advancing spin-wave-based computing and data processing applications.

\newpage
\bibliography{bibliography}
\bibliographystyle{ieeetr}

\end{document}


\title{Supplementary Material: Spatial Control of Hybridization Induced Spin-Wave Transmission Stop Band}

\author[1]{Franz Vilsmeier\footnote{franz.vilsmeier@tum.de}}

\author[1]{Christian Riedel}
\author[1]{Christian H. Back}
\affil[1]{\textit{Fakultät für Physik, Technische Universität München, Garching, Germany}}
\date{March 23, 2024}

\maketitle

\section{Time-Resolved Magneto-Optic Kerr Effect Microscopy}

As a source of illumination, a mode-locked Ti:Sa laser with a centre wavelength of 800\,nm and a pulse width of around 150\,fs is used. The pulse trains are applied at a fixed repetition rate of 80\,MHz. Subsequently, we fix the polarization plane of the laser and focus it onto the sample through an objective lens with a numerical aperture of 0.7, giving a maximum resolution of $\sim$0.6\,$\upmu$m. Upon reflection at a magnetic surface, the polarization changes due to the polar magneto-optical Kerr effect. Here, the change of polarization rotation is directly proportional to the change in the dynamic out-of-plane magnetization component. A Wollaston prism splits the reflected signal into two beams with orthogonal polarization components, which are detected by two photodiodes. The difference between the two photodiode signals then gives a direct representation of the change in magnetization - the Kerr signal -, and the sum of the two is proportional to the sample's reflectivity. Since the sample is mounted onto a piezo stage, the relative laser focus position can be spatially scanned in the sample plane. Hence, Kerr image and topography are obtained. 

In addition, during the acquisition of a Kerr image, the relative phase relation between the applied rf-frequency in the GHz regime and the laser repetition rate is fixed. This requires the driving field frequency to always be a multiple of the laser repetition rate. As a result of the constant phase and the short laser pulses (much shorter than one period of the excitation), we can directly access the dynamic out-of-plane magnetization component and observe the propagation of spin-waves excited by our antenna structure.

\begin{figure}[h]
    \includegraphics[width=\textwidth]{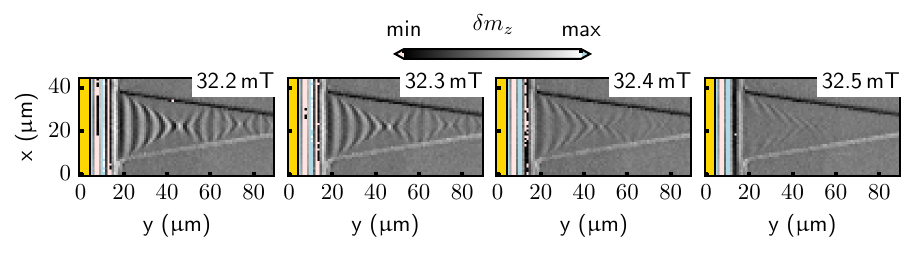} 
    \caption{Kerr images recorded at $f=2.8$\,GHz at different fields coming from below the stop band. Caustic-like beams are reflected back and forth at the edges and fade away along the trapezoid with increasing field.}
    \label{fig:SupplementaryMaterial_Figure1}
\end{figure}

Fig.~\ref{fig:SupplementaryMaterial_Figure1} shows some spatial Kerr maps approaching the full film hybridization condition from lower fields. Caustic-like beams emerge, which are damped along the trapezoid with increasing field. 

\section{Effect of Waveguide Width on Hybridization}

Following the analytical model by Kalinikos and Slavin~\cite{Kalinikos1986}, the full film dispersion relation in the presence of an in-plane external field  with totally unpinned surface states is given by
\begin{equation}
    \label{eq:dispersion}
    \omega_n^2= \left(\omega_\mathrm{H} + l_\mathrm{ex}^2k_n^2\omega_\mathrm{M}\right)\left(\omega_\mathrm{H} + l_\mathrm{ex}^2k_n^2\omega_\mathrm{M} + \omega_\mathrm{M}F_{nn}\right), 
\end{equation}
where
\begin{equation}
    \omega_\mathrm{H}=\gamma\mu_0 H,
\end{equation}
\begin{equation}
    \omega_\mathrm{M}=\gamma\mu_0 M_\mathrm{S},
\end{equation}
\begin{equation}
    F_{nn}=P_{nn} + \left(1-P_{nn}\left(1+\cos^2\varphi\right)+\omega_\mathrm{M}\frac{P_{nn}\left(1-P_{nn}\right)\sin^2\varphi}{\omega_\mathrm{H}+l_\mathrm{ex}^2k_n^2\omega_\mathrm{M}}\right),
\end{equation}
and
\begin{equation}
\begin{aligned}
    P_{nn}= & \frac{k^2}{k_n^2}-\frac{k^4}{k_n^4}F_n\frac{1}{1+\delta_{0n}}, \\
    F_n= & \frac{2}{kL}\left(1-(-1)^n\mathrm{e}^{-kL}\right).
\end{aligned}
\end{equation}
Furthermore, $n=0,1,2,...$ denote the eigenmode orders across the film thickness $L$, $k_n=\sqrt{k^2+\left(\frac{n\pi}{L}\right)^2}$, and $\varphi$ describes the angle between $\textbf{k}$ and $\textbf{M}$ (so for $\textbf{k}\perp\textbf{M}$, $\varphi=\frac{\pi}{2}$).

Considering a spin wave waveguide of finite width $w$, an additional quantization across the waveguide width is introduced and the dispersion relation can be represented using equ.~(\ref{eq:dispersion}) by letting $k\to\sqrt{k^2+\left(\frac{m\pi}{w}\right)^2}$ and $\varphi\to\varphi-\arctan\left(\frac{m\pi}{kw}\right)$~\cite{Demidov2015,Brcher2017,Chumak2019}. Here, $m=0,1,2,...$ denote the eigenmode orders across the width, and $k$ denotes the wavenumber along the waveguide. In the specific case of a tangentially magnetized waveguide in the DE-geometry ($\textbf{k}\perp\textbf{M}$), demagnetization also has to be taken into account and the non-uniform effective field $\mu_0H_\mathrm{eff}$ needs to be considered in the dispersion relation in place of the externally applied field. In the case of spin wave propagation in the center, a uniform field is assumed, but an effective waveguide width $w_\mathrm{eff}$ is introduced to account for the strong reduction in the effective field at the edges. The effective width can be defined in different ways. Here, we follow the definition by Chumak~\cite{Chumak2019}, where $w_\mathrm{eff}$ is given by the distance of points across the width where the effective field is reduced by 10\%. i.e., to the value $0.9\cdot\mu_0H_\mathrm{eff}^{\mathrm{max}}$. 

\begin{figure}[h]
    \includegraphics[width=\textwidth]{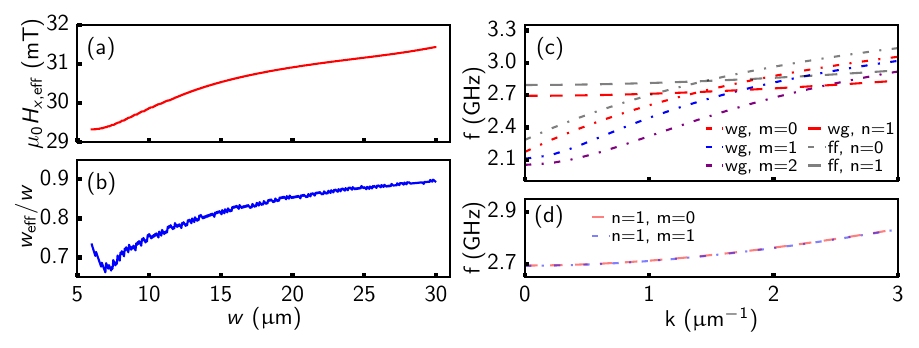}
    \caption{Effect of width modes on dispersion relation. (a) Effective field from micromagnetic simulations for the center mode as a function of trapezoid width $w$. (b) Ratio of extracted effective width $w_\mathrm{eff}$ and actual trapezoid width $w$. (c) Dispersion relations of the waveguide (wg) modes considering effective field and effective width for $w=6\,\upmu$m. Modes with n=0 and m=0 (blue dash-dotted line), n=0 and m=1 (red dash-dotted line), n=0 and m=2 (purple dash-dotted line), and n=1 and m=0 (red dashed line) are shown. The full film (ff) modes with n=0 and n=1 are also depicted (grey lines) for comparison. (d) Waveguide modes with n=1 and m=0, and n=1 and m=1. The width modification does not significantly affect the first-order PSSW.}
    \label{fig:SupplementaryMaterial_Figure7}
\end{figure}

From micromagnetic simulations~\cite{MuMax2014}, the effective field within the trapezoid geometry at an externally applied field of 32\,mT (close to the full film hybridization field at 2.8\,GHz) was determined. From the field distribution, the effective field and effective width for the center mode were extracted as a function of trapezoid width $w$. The respective results are depicted in Figs~\ref{fig:SupplementaryMaterial_Figure7}(a)-(b). At the smallest width, the effective field is reduced by almost 3\,mT with respect to the applied field. The effective width is maximally reduced to about 65\,\% of the actual waveguide width, allowing for a rather wide region of uniform field and mode propagation across the width.

For $w=6\,\upmu$m, the resulting dispersion relations for a waveguide with several width modes (m=0, m=1, m=2) and thickness modes (n=0, n=1) are displayed in Fig.~\ref{fig:SupplementaryMaterial_Figure7}(c). Note that for the waveguide case, only the m=0 mode of the first-order (n=1) PSSW mode is shown, as the width quantization doesn't notably affect the higher-order PSSWs (see Fig.~\ref{fig:SupplementaryMaterial_Figure7}(d)). The reduced effective field generally shifts the dispersion relation towards lower frequencies compared to the full film case (grey lines). The higher-order width modes (m=1, m=2) also display lower frequencies in the dipolar regime than the m=0 mode. More importantly, however, the n=0 and n=1 thickness modes still intersect in the dipolar regime, facilitating a hybridization and corresponding stop band. From this, we conclude that the main effect of the width modulation on the hybridization is the reduced effective field and the resulting shift in the hybridization condition. This is especially the case for the m=0 width mode, which should be dominant in the trapezoid structure due to the transmission of spin waves from the full film into the tapered waveguide.

\section{Broadband Spin-Wave Spectroscopy}

\begin{figure}[h]
    \includegraphics[width=0.75\textwidth]{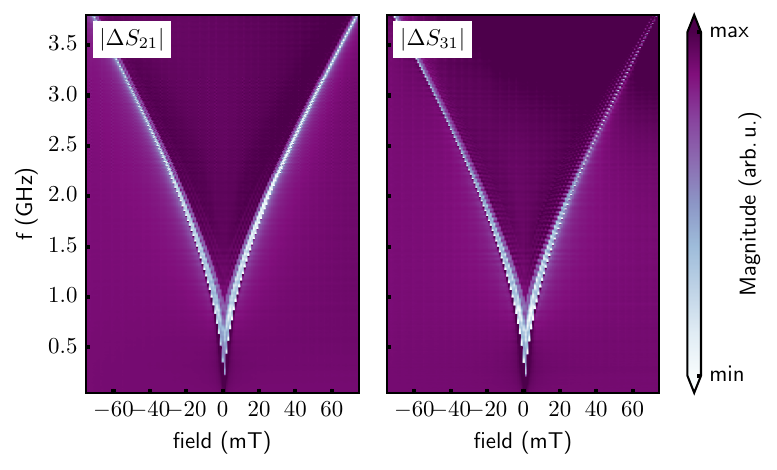}
    \caption{Broadband spin-wave spectroscopy spectra. Modes close to FMR are very prominent in the transmission spectra $|\Delta S_\mathrm{21}|$ and $|\Delta S_\mathrm{31}|$.}
    \label{fig:SupplementaryMaterial_Figure3}
\end{figure}

A four-port vector network analyzer (Agilent N5222A) was used for the broadband spin-wave spectroscopy. All measurements were conducted at a microwave power of 3\,dBm, and the real and imaginary parts of the complex scattering parameters $S_\mathrm{21}$ and $S_\mathrm{31}$ were recorded. A frequency sweep method was applied at different external magnetic field values for the transmission spectra. The magnetic field's strength was changed stepwise (5\,mT steps) from high to low field. To improve contrast, a high-field subtraction method was applied. Reference data $S_\mathrm{21,ref}$ and $S_\mathrm{31,ref}$ taken at 200\,mT was recorded. The presented spectra were then obtained by subtracting the absolute value of the reference data from the absolute value of the scattering parameters, i.e., $|\Delta S_\mathrm{21}|=|S_\mathrm{21}|-|S_\mathrm{21,ref}|$ and $|\Delta S_\mathrm{31}|=|S_\mathrm{31}|-|S_\mathrm{31,ref}|$. Exemplary recorded spectra are shown in Fig.~\ref{fig:SupplementaryMaterial_Figure3} where modes close or at the FMR are prominent. In the CW mode measurements, no reference was taken.

\section{Edge Mode Due to Imperfect Fabrication}

\begin{figure}[t]
    \includegraphics[width=\textwidth]{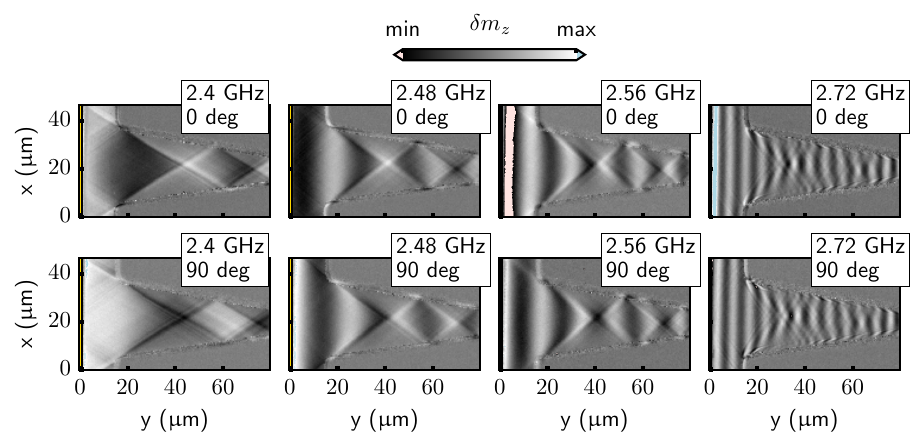}
    \caption{TR-MOKE measurements for different frequencies and phases between microwave excitation and laser pulses at an external field of 33.5\,mT. In the vicinity where the caustic-like beams scatter from the edges, an additional mode profile is observed. We note that the antisymmetric beam directions stem from a slight mismatch of the external field angle with respect to the DE-geometry.}
    \label{fig:SupplementaryMaterial_Figure4}
\end{figure}

Fig.~\ref{fig:SupplementaryMaterial_Figure4} shows TR-MOKE measurements at different frequencies and relative phases between the microwave excitation and laser pulses at an applied field of 33.5\,mT. Apart from the propagation inside the trapezoid, Kerr images where intense caustic-like beams are detected (2.4\,GHz, 2.48\,GHz, 2.56\,GHz) also exhibit a magnetic contrast right at the edges of the patterned YIG. In close vicinity to the points where the beams are reflected from the trapezoid edges, a localized mode profile outside the previous propagation region in $x$-direction is visible. However, no such distinct feature seems to occur when DE-like modes become dominant in the profile (2.72\,GHz).

\begin{figure}[hbt!]
    \includegraphics[width=\textwidth]{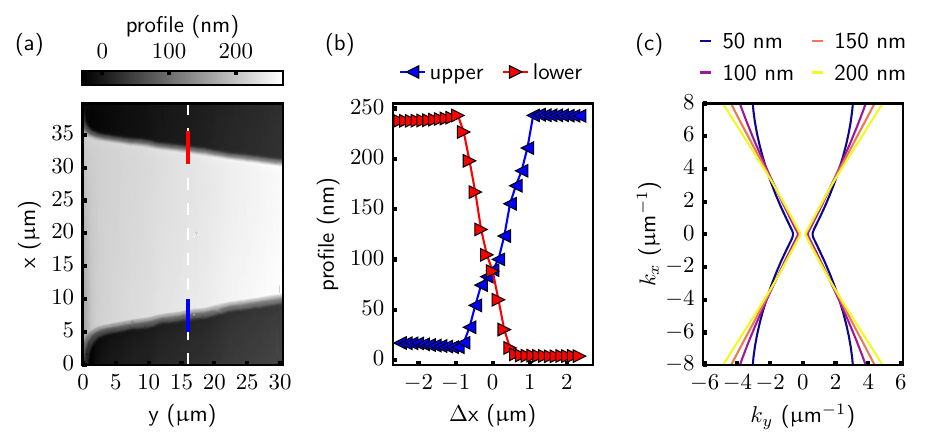}
    \caption{(a) AFM profile of patterned YIG structure. (b) Linescans taken across the edges highlighted in blue and red in (a). A gradual transition from the YIG to the GGG is observed. (c) Iso-frequency curves for several film thicknesses at 2.48\,GHz and 33.5\,mT.}
    \label{fig:SupplementaryMaterial_Figure5}
\end{figure}

Fig.~\ref{fig:SupplementaryMaterial_Figure5}(a) depicts an AFM image of the patterned YIG trapezoid. Linescans across the edges (Fig.~\ref{fig:SupplementaryMaterial_Figure5}(b)) reveal that the transition from YIG to GGG substrate is not sharp, but rather gradual along a distance of about 1-2\,$\upmu$m. This is attributed to imperfections in the  fabrication process.

The boundary regions, from which the caustic-like beams scatter, serve as secondary point-like excitation sources with a finite size of the order of the beam's width, as noted in previous works~\cite{Chumak2019,Schneider2010}. Consequently, spin wave modes may be excited within the transitional region where the thickness of YIG decreases. Examining the iso-frequency curves at 2.48\,GHz and 33.5\,mT (see Fig.~\ref{fig:SupplementaryMaterial_Figure5} (c)) reveals that the modes potentially excited fall within our resolution limits across a considerable range of film thicknesses.

\section{Some Properties of Hybridization-Induced Stop Band}

This section offers a brief overview of some characteristics of the anticrossing in full YIG films. All micromagnetic simulations were executed utilizing the \textsc{TetraX}~\cite{TetraX2022} software package.

To provide a qualitative understanding of the hybridization's coupling strength, we introduce the quantity $\Delta f$ as the minimal gap between the upper and lower band determined by micromagnetic simulations. We note that here, we only consider the strength of hybridization in the frequency domain, as this needed significantly less computing time. Furthermore, we inferred the wave vector of hybridization $k_\mathrm{hyb}$ by the intercept of the n=0 and n=1 modes according to the model by Kalinikos and Slavin~\cite{Kalinikos1986}.

\begin{figure}[h]
    \includegraphics[width=\textwidth]{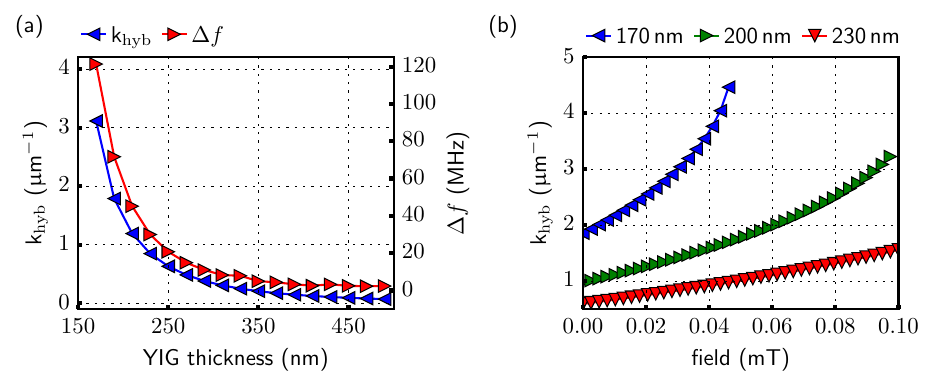}
    \caption{Some properties of hybridization. (a) Trend of hybridization wave number $k_\mathrm{hyb}$ and hybridization strength  with increasing film thickness at external field of 32\,mT. Both parameters decrease with increasing thickness. (b) Dependence of $k_\mathrm{hyb}$ on the external field for different film thicknesses. $k_\mathrm{hyb}$ can be increased to some extent by the external field strength.}
    \label{fig:SupplementaryMaterial_Figure6}
\end{figure}

In Fig.~\ref{fig:SupplementaryMaterial_Figure6}(a), the relationship between film thickness and both $\Delta f$ and $k_\mathrm{hyb}$ is illustrated at an external field strength of 32\,mT. As film thickness increases, both these parameters exhibit a decreasing trend. Notably, below a thickness of 170\,nm, the formation of an anticrossing appears to be absent. Here, the increased separation between n=0 and n=1 prevents hybridization from occurring. We also note that potential crossing is only possible in the dipolar regime as modes follow $k^2$-dependence in the exchange regime. Fig.~\ref{fig:SupplementaryMaterial_Figure6}(b) depicts the influence of the external field on $k_\mathrm{hyb}$. Higher external field strengths result in higher wave numbers. Furthermore, the external field strength also affects the existence of frequency degeneracy. At higher fields, the n=0 mode becomes flatter and no longer intersects with the n=1 mode. In summary, the manipulation of external field strength and sample thickness allows for the adjustment of hybridization properties, facilitating higher wave number values and stronger coupling within a specific range.

\bibliography{supplement_bibliography}
\bibliographystyle{ieeetr}